\documentclass[a4paper,fleqn,usenatbib]{mnras}

\usepackage[T1]{fontenc}
\usepackage{ae,aecompl}

\usepackage{graphicx}	
\usepackage{amsmath}	
\usepackage{amssymb}	

\title[Binary star detectability in {\it Kepler} data]{Binary star detectability in {\it Kepler} data from phase modulation of different types of oscillations}
\author[Compton et al.]{D. L. Compton$^{1,2}$\thanks{E-mail:
d.compton@physics.usyd.edu.au}, T. R. Bedding$^{1,2}$, S. J. Murphy$^{1,2}$, D. Stello$^{1,2}$\\
$^{1}$ Sydney Institute for Astronomy (SIfA), School of Physics, University of Sydney, NSW 2006, Australia \\
$^{2}$ Stellar Astrophysics Centre, Department of Physics and Astronomy, Aarhus University, Ny Munkegade 120, DK-8000 Aarhus C, Denmark}

\begin{document}

\date{Accepted 2016 May 4. Received 2016 May 1; in original form 2016 March 22}

\pagerange{\pageref{firstpage}--\pageref{lastpage}} \pubyear{2015}

\maketitle

\label{firstpage}

\begin{abstract}
Detecting binary stars in photometric time series is traditionally done by measuring eclipses. This requires the orbital plane to be aligned with the observer. A new method without that requirement uses stellar oscillations to measure delays in the light arrival time and has been successfully applied to $\delta$~Scuti stars. However, application to other types of stars has not been explored. To investigate this we simulated light curves with a range of input parameters. We find a correlation between the signal-to-noise of the pulsation modes and the time delay required to detect binary motion. The detectability of the binarity in the simulations and in real {\it Kepler} data shows strong agreement, hence, we describe the factors that have prevented this method from discovering binary companions to stars belonging to various classes of pulsating stars.
\end{abstract}

\begin{keywords}
stars : oscillations - stars : variables - stars : binaries - techniques : radial velocities
\end{keywords}

\section{Introduction}

The primary science goal of {\it Kepler} was to find stars with Earth-like exoplanet companions by observing transits in the photometric time series \citep[][]{koch10a,borucki10}. Large numbers of eclipsing binaries were also discovered \citep{prsa11,kirk16}, some of which show stellar oscillations \citep[e.g.][]{southworth15}. These oscillations can also be used to find binaries without transits or eclipses, provided the pulsation modes are stable enough to act like a `clock'. That is the subject of this paper.

As a star and its companion orbit each other, the light travel time from the host star to the observer will vary. The variation is a periodic function that is related to the integral of the radial velocity variation. \citet{telting12} used the differential arrival time of pulsation modes to confirm the presence of a companion to an sdB star. \citet{murphy14a} developed the phase modulation (PM) method, using changes in the pulsation phases of $\delta$ Sct stars to find binary companions. This method complements the frequency modulation (FM) method by~\citet{shibahashi12}, wherein the binary motion modulates the oscillation frequencies, causing multiplets in the Fourier transform. The FM method is suitable for data sets that are much longer than the orbital period of the binary. Hence, for the wide orbits the frequency splitting can approach the frequency resolution of the pulsation spectrum. In contrast, the PM method favours wider orbits because the light travel time across the orbit is larger. The PM method has the benefit of providing a visualisation of the orbit by tracking the time delays as a function of orbital phase. It also allows the signals from different pulsation modes to be combied straightforwardly. This method has been used on known $\delta$~Scuti binaries in the {\it Kepler} field \citep[e.g.][]{balona14a}. Additionally, many new binary systems have been discovered using the PM method \citep[e.g.][]{murphy14a}. Several giant planets have been discovered around pulsation sdB stars using timing variations \cite[e.g.][]{silvotti07,lee09,qian09,geier09}.

The goal of this paper is to apply the PM method to other types of pulsating stars to detect the presence of a binary companion. We start by introducing the PM method and apply it to artificial light curves (Section~\ref{sec:meth}). We adopt a Monte Carlo method by mass-producing light curves and extracting relevant parameters from the time-delay spectrum. The resulting distributions are analysed to diagnose detectability and then compared with real {\it Kepler} binaries in Section~\ref{sec:res}. Finally the implications of this research are discussed in Section~\ref{sec:disc}.

\section{Method and simulations}
\label{sec:meth}

\subsection{Phase measurements and time delays}
\label{sec:extract}

The PM method involves measuring the phase,~$\Phi(t)$, of a pulsation mode over time,~$t$. We generalise the light curve as a periodic function,~$f(t)$. The phase of a pulsation mode with frequency~$\nu$ can be calculated using the Fourier transform over a time interval~$\delta t$, which is

\begin{equation}
\label{equ:fourier}
{\rm F}(t;\nu,\delta t) = \int^{t+\delta t/2}_{t-\delta t/2} f(t') e^{-2 \pi i \nu t'} dt'.
\end{equation}
The argument of the complex quantity~$F$ in Equation~\ref{equ:fourier} gives the phase:
\begin{equation}
\label{equ:fphase}
\Phi(t;\nu) = \tan^{-1}\left(\frac{{\rm Im}({\rm F}(t;\nu,\delta t))}{{\rm Real}({\rm F}(t;\nu,\delta t))}\right).
\end{equation}
The length of the intervals used by \citet{murphy14a} was $\delta t = 10$ days.

The phase of a pulsation signal is sensitive to changes in distance to the source. Extracting the pulsation phase for each time interval produces a series of phases that are modulated by orbital motion, allowing us the measure the binary period,~$P_{\rm orb}$, and projected semi-major axis $a_1 \sin{(i)}$, where $i$ is the inclination of the orbital plane with respect to the observer. 

The phase variations can be converted to time delays by dividing by the angular frequency of the pulsation mode:
\begin{equation}
\label{equ:dtau}
\tau(t) = \frac{\Delta \Phi(t)}{2 \pi \nu}.
\end{equation}
Here, $\Delta \Phi = \Phi(t) - \left<\Phi(t)\right>$, which sets the mean of the time delays to be zero. 

\begin{figure}
	\centering
	\includegraphics[width=0.99\columnwidth]{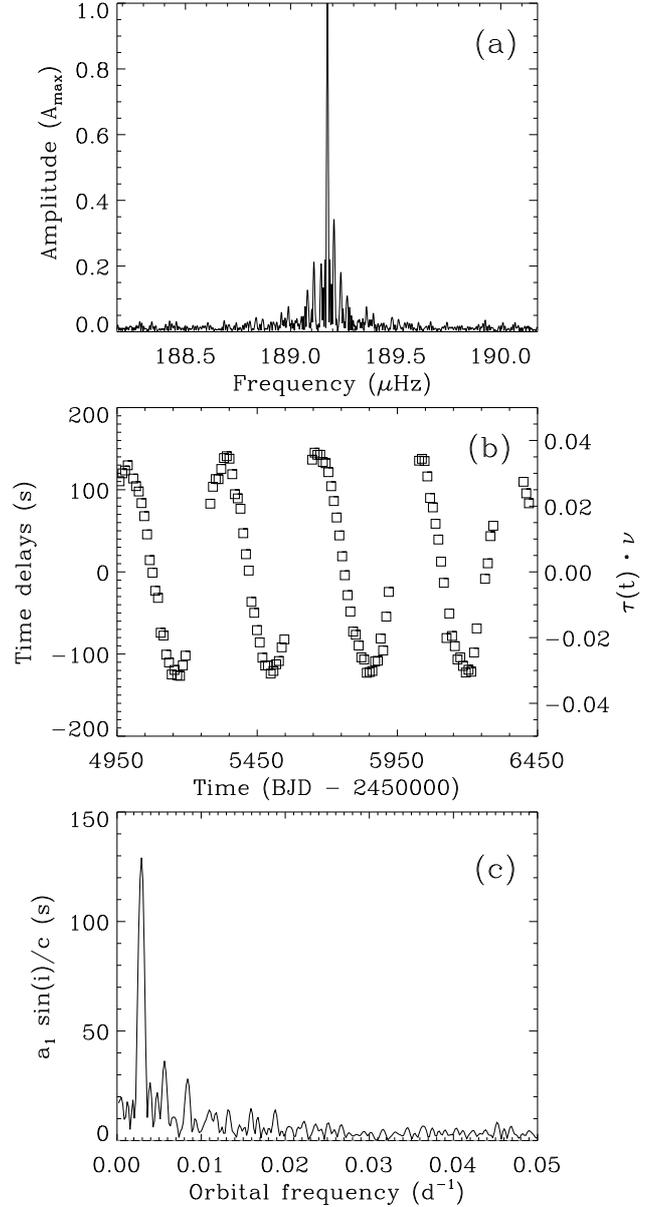}
	\caption{Application of the phase modulation method to the {\it Kepler} $\delta$ Sct binary star KIC\,11754974. (a) The close-up view of a single mode at $189.1\,\mu{\rm Hz}$ in the Fourier transform in the photometric time series. (b) time delays calculated from $10\,{\rm d}$ sub-series at the frequency of the chosen mode. (c) Fourier transform of the time delays (time-delay spectrum). The peak at $0.0029\,{\rm d}^{-1}$ in the time-delay spectrum is caused by the binary motion of the $\delta$ Sct star.}
	\label{fig:tsps}
\end{figure}

Figure~\ref{fig:tsps} shows the main steps of the PM method for a known {\it Kepler} binary $\delta$~Sct star KIC\,11754974~\citep[][]{murphy13b}: (a) the Fourier transform of the time series around a prominent oscillation mode, (b) time-delay series, and (c) time-delay spectrum. Binary motion produces a peak in the Fourier transform of the time delays (henceforth time-delay spectrum) at the orbital frequency,~$f_{\rm orb}=1/P_{\rm orb}$. The amplitude of this peak is the projected light travel time across the orbit, which is $a_1 \sin{(i)}$ divided by the speed of light~$c$. A more complete set of orbital parameters can be extracted from fitting directly to the time delays \citep[see][]{murphy15a}.

\subsection[]{Simulated light curves}
\label{sec:smts}

To evaluate the performance of the PM method, we generated monoperiodic {\it Kepler} time series, which simulate the binary motion. The simulated light curves were a linear combination of two components: a single oscillation mode and a noise term. Binarity was simulated by adding a time-dependent phase shift to the pulsation. 

The flux variation due to a single mode in the time series is given by
\begin{equation}
\label{equ:smode}
f(t_n; A,\nu,\phi) = A \cos{(2 \pi \nu (t_n + \tau(t_n)) + \phi )},
\end{equation}
where $A$ and $\nu$ are the amplitude and frequency of the mode, and $\phi$ is the phase relative to an arbitrary fixed starting point. The time stamps,~$t_n$, for the simulations were based on the {\it Kepler} long cadence sampling ($\Delta t = 29.4{\rm min}$) using all four years of available data. The time-dependent function $\tau(t_n)$ describes the time delays induced by the binary motion. To simplify our analysis we only considered circular orbits, i.e. eccentricity $e$ equals zero. \citet{shibahashi12} expressed the phase modulation function as
\begin{equation}
\label{equ:phasemod}
\tau(t_n;\tau_{\rm max},P_{\rm orb},\psi) = \tau_{\rm max} \sin{\left( \frac{2 \pi t_n}{P_{\rm orb}} \right) + \psi},
\end{equation}
where $\tau_{\rm max}$~is the amplitude of the time-delay variations, $P_{\rm orb}$~is the orbital period, and $\psi$~describes the phase of the orbit. For a circular orbit, $\tau_{\rm max}$ only depends on the projected semi-major axis. In the absence of noise, the maximum time delay is equivalent to the projected light travel time across the orbit, i.e. $\tau_{\rm max}=a_1 \sin{(i)}/c$. In this paper $\tau_{\rm max}$ will be used as the input maximum time delay, whereas $a_1 \sin{(i)}/c$ is the empirical maximum time delay extracted from the simulated time series using the PM method. 

We simulated the white noise using the equation
\begin{equation}
W(t_n;\sigma) = X_{t_n} (\sigma_{\rm rms}),
\end{equation}
where $X_{t_n}$~is a random number taken from a Gaussian distribution with a mean of zero and a standard deviation of $\sigma_{\rm rms}$.

The relationship between the scatter in the time series,~$\sigma_{\rm rms}$, and the mean noise level in the amplitude spectrum,~$\sigma_{\rm amp}$, is
\begin{equation}
\label{equ:sigamp}
\sigma_{\rm amp} = \sqrt{\frac{\pi}{N}}\sigma_{\rm rms}.
\end{equation}
Here, $N$ is the number of data points in the time series \citep[][]{kjeldsen95}. The relationship between $A$ and $\sigma_{\rm rms}$ gives the signal-to-noise ratio of the oscillation mode,
\begin{equation}
\label{equ:snamp}
{\rm S/N} = \frac{A}{\sigma_{\rm amp}} = \frac{A}{\sqrt{\frac{\pi}{N}}\sigma_{\rm rms}}.
\end{equation}
The signal-to-noise ratio is a convenient quantity because it combines the oscillation amplitude and noise into one scalar quantity and it can be measured straight-forwardly from the Fourier transform of the light curve. 

The uncertainty of the pulsation phase measurement depends on this signal-to-noise. For a given binary orbit, Equation~\ref{equ:dtau} indicates that the phase modulation of a particular pulsation mode is proportional to the mode frequency. It follows that the uncertainties of the phases are lower for higher pulsation frequencies. 

The randomness of our simulations gives a distribution of measured maximum time delays,~$a_i \sin{(i)}/c$, and orbital periods,~$P_{\rm obs}$. Non-varying asteroseismic and binary parameters were marginalised by setting them as constant across all simulations, as shown in Table~\ref{tab:param}.

\begin{table}
\centering
\caption{Set of fixed parameters used in the simulations. $Y$ is a uniform random-number-generating-function between 0 and 1. Each simulation used a different random seed.}
\label{tab:param}
\begin{tabular}{@{}lcl}
\hline
Parameter & Description & Simulation Value \\
\hline
$P_{\rm orb}$ & Orbital period & $100\,{\rm d}$  \\
$e$ 		  & Orbital eccentricity & $0$  \\
$\psi$        & Orbital phase & $ 2 \pi Y_i$\\
$N_{\rm sub}$ & Number of sub-series & $150$  \\
$A $          & Pulsation amplitude & $1.0$ (arbitrary units)  \\
$\nu_{\rm sim}$       & Pulsation frequency & $250\,\mu{\rm Hz}$  \\
$\phi$        & Pulsation phase & $ 2 \pi Y_j$\\

\hline
\end{tabular}

\end{table}

The amplitude,~$A$, was kept constant and the noise level,~$\sigma_{\rm rms}$, was adjusted to control the signal-to-noise of the mode. We split each time series into $N_{\rm sub}=150$ sub-series, which corresponds to approximately ten days for most {\it Kepler} light curves. If the effective length of a sub-series was less than eight days, it was discarded. Ten days is short enough to sample the orbit for any orbital period above 20~d and long enough to resolve individual modes with a minimum frequency separation of approximately $3 \mu{\rm Hz}$ \citep[see][]{jcd08b}.

In general, high-frequency pulsations and longer orbital periods are advantageous. However, upper limits exist for both these quantities. A full sample of the orbit throughout the time series (i.e. $P_{\rm orb} \lesssim 1000{\rm d}$) is required to ensure that the time-delay variations are periodic and caused by an orbiting companion. The pulsation frequency is limited by the long-cadence sampling, which has a Nyquist frequency of $283.2 \mu{\rm Hz}$. Therefore, care must be taken to avoid aliases \citep[e.g.][]{murphy13a}.

\subsection{Setting limits on the detection of binarity}
\label{sec:nobino}

To determine the limits of binary detectability we generated simulated {\it Kepler} time series with a grid of input parameters. We initially simulated time series without binarity to understand the influence of the pulsation mode signal-to-noise on the time-delay spectrum. We simulated 1000 light curves for each value of signal-to-noise in a grid spanning $5 < {\rm S/N} < 2000$. The PM method was applied to each white-noise time series to extract the height of the highest noise peak from the time-delay spectrum. For a given signal-to-noise ratio, the absolute phase uncertainty caused by the white noise is the same across different modes of varying frequency. This means that the variance of this extracted maximum time delay multiplied by the pulsation mode frequency is the same for modes of identical signal-to-noise ratios. Therefore, we compared the phase modulation in units of number of pulsation periods across stars with varying white noise coefficients, i.e. time time delays multipled by the speed of light: $\tau_{\rm max} \cdot \nu$ or $a_1 \sin{(i)} / c \cdot \nu$.

\begin{figure*}
	\centering
	\includegraphics[width=1.99\columnwidth]{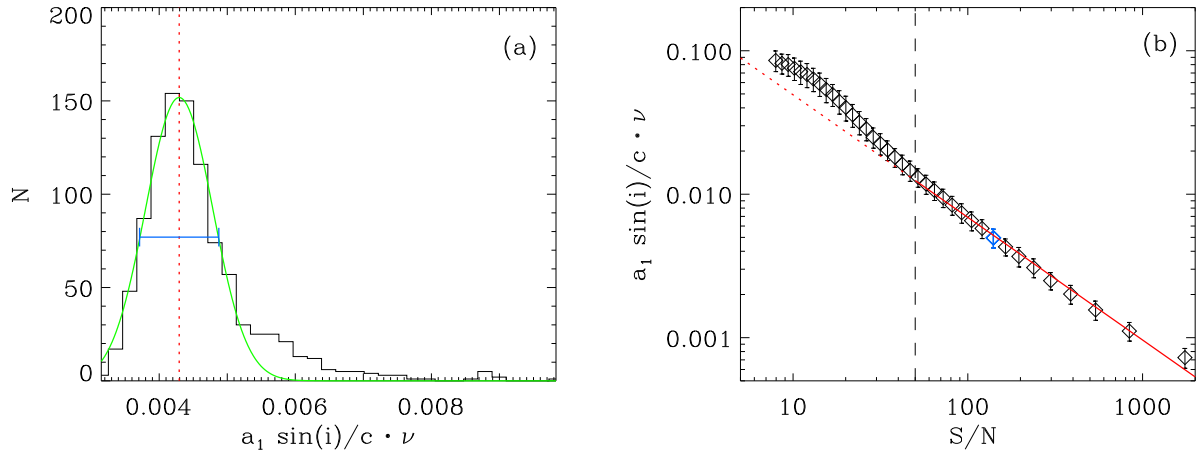}
	\caption{(a) A single distribution (${{\rm S/N} \approx 140}$) of the maximum noise-peak in the time delay spectrum using $N=1000$ simulated time series of the same star with no binary component. The distribution was fitted with a Gaussian (green line). The mean and standard deviation of the Gaussian is represented by the dotted red line and the blue bar, respectively. (b) The diamonds and attached error bars represent the mean and standard deviation, respectively, of each distribution as a function of signal-to-noise. The blue diamond corresponds to the distribution in (a). The solid red line is a power-law fit to the points with  ${\rm S/N} > 50$ (dashed black line), and the dotted red line is its extrapolation.}
	\label{fig:avmaxtau}
\end{figure*}

An example of the distribution of maximum time delay from 1000 simulations of purely white noise time series is shown in Fig.~\ref{fig:avmaxtau}a. A Gaussian fit to the histogram gives the typical maximum time delay caused by the white noise scatter. The maximum time delay of a binary must exceed this value for a given signal-to-noise ratio to be considered detectable, i.e. above the points in Fig.~\ref{fig:avmaxtau}b. At low ${\rm S/N}$ we observed excessive phase-wrapping in the time-delay sub-series, that is the point-to-point scatter in Fig.~\ref{fig:tsps}b exceeded the co-domain of Equation~\ref{equ:fphase} ($|\Phi(t)| > \pi$). We fitted a power-law to the points above a signal-to-noise ratio of 50, which we considered to be a soft lower bound on the binary detection threshold for a monoperiodic star due to the phase wrapping.

\subsection{Simulations of binary motion}

We extended our analysis by adding phase modulation ($\tau_{\rm max} > 0$ in Equation~\ref{equ:phasemod}) to simulate binary motion. We used a similar Monte Carlo analysis to find the uncertainty of the extracted maximum time delay as a function of $\tau_{\rm max}$ and pulsation mode ${\rm S/N}$.  Arrays of ${\rm S/N}$ and $\tau_{\rm max}$ were logarithmically spaced to make a 40x40 grid of input parameters. For each grid point another 1000 monoperiodic time series were constructed with different random seeds. Using the PM method, the maximum time delay,~${a_1 \sin{(i)}/c}$, was extracted from the time-delay spectra. Note that we scaled the time delays to a phase unit,~$a_1 \sin{(i)} / c \cdot \nu$, to allow easy comparison between stars in different binary systems or having different oscillation frequencies. 

\begin{figure}
	\centering
	\includegraphics[width=0.99\columnwidth]{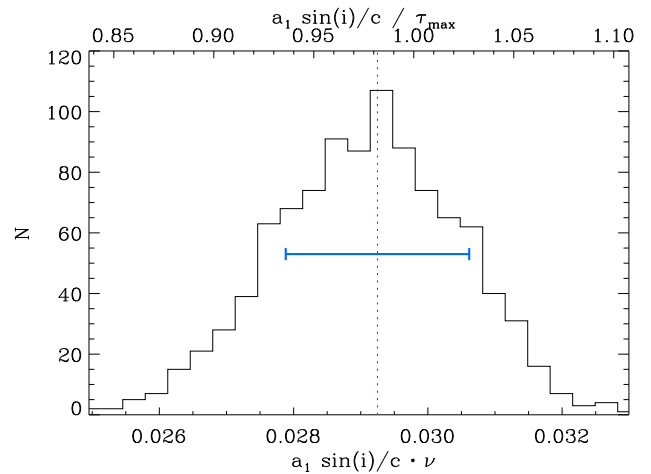}
	\caption{The distribution of maximum time delays from 1000 simulations with input parameters of ${\tau_{\rm max}~\approx 120~{\rm s}}$ and ${{\rm S/N} \approx 140}$. The red dashed line is the mean of the distribution. The one standard deviation uncertainties are represented by the blue bar. The top axis denotes the ratio of the highest peak extracted from the time-delay spectra,~$a_1 \sin{(i)} / c$, and the maximum time delay injected into the simulated,~$\tau_{\rm max}$.}
	\label{fig:distoex}
\end{figure}

We then looked at the distribution of maximum time delays of each of the sets of 1000 simulated time series in the grid. These distributions were found to be Gaussian-like, as in Fig.~\ref{fig:distoex}, as long as the time delay due to binary was consistently above the maximum noise-peak. In contrast, the maximum time delay distribution of a set of noisy time series was found to be spread out and non-Gaussian because more false peaks in the time-delay spectra were extracted. The mean of the maximum time delay distribution in Fig.~\ref{fig:distoex} is less than the input maximum time delay. This is caused by an undersampling of the orbit. Murphy et al. (in prep) will give a full characterisation of the effect, but with our sampling of 10 sub-series per orbit there is little impact on our results.

\begin{figure*}
	\centering
	\includegraphics[width=1.99\columnwidth]{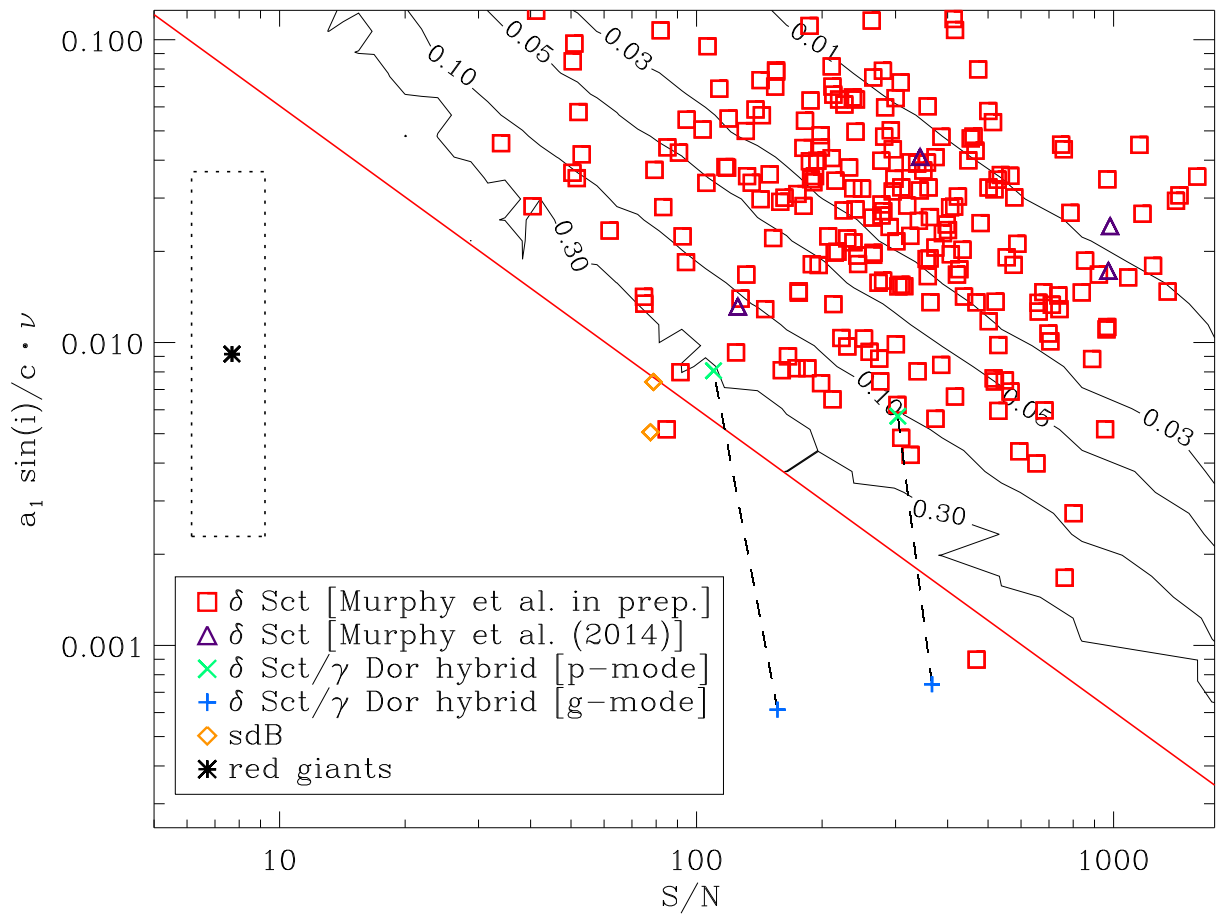}
	\caption{The contours define lines of constant relative uncertainty in the time delays calculated from a sample of 1000 simulated light curves as a function of signal-to-noise ratio,~${\rm S/N}$ of the pulsation mode, and injected maximum time delay,~$\tau_{\rm max}$. The red line is the power-law fit to the average maximum noise peak, shown in Fig.~\ref{fig:avmaxtau}b. Each symbol represents the strongest pulsation mode of a known {\it Kepler} binary star. The red squares and purple triangles are $\delta$~Sct p~modes (by ~\citet{murphy14a}). Orange diamonds are sdB stars measured by~\citet{telting12,telting14} (KIC\,7668647 and KIC\,11558725). The dotted-lined box represents the area where the most ideal red giants lie \citep[][]{beck14,gaulme14}. The green crosses and blue plusses are $\delta$ Sct/$\gamma$ Dor hybrids, p mode and g mode, respectively (see \citet[][]{vanreeth15} for KIC\,3952623 and \citet[][]{keen15} for KIC\,10080943).}
	\label{fig:ex}
\end{figure*}

The grid of distributions are compiled into Fig.~\ref{fig:ex} using contours of constant relative uncertainty. The relative uncertainty was calculated by dividing the standard deviation of each distribution by the injected maximum time delay,~$\tau_{\rm max}$. The contour lines follow a power-law for large ${\rm S/N}$ and time-delay amplitude, where binarity is most easily detectable. Conversely, the contours become irregular at larger uncertainties and the binary is harder to detect. The large uncertainties correspond to a parameter space where the noise in the time-delay spectrum consistently exceeds $\tau_{\rm max}$ (red line in Fig.~\ref{fig:ex} and Fig.~\ref{fig:avmaxtau}b). We concluded that the binary peak in the time-delay spectrum could not be reliably detected if the relative uncertainty was much greater than 30\%. Therefore, contours greater than this are not shown in Fig.~\ref{fig:ex}. We should keep in mind that our simulation only included a single pulsation mode in each time series, and that we should expect to do somewhat better in multiperiodic stars.

\section{Comparison with observed data}
\label{sec:res}

\begin{figure*}
	\centering
	\includegraphics[width=1.99\columnwidth]{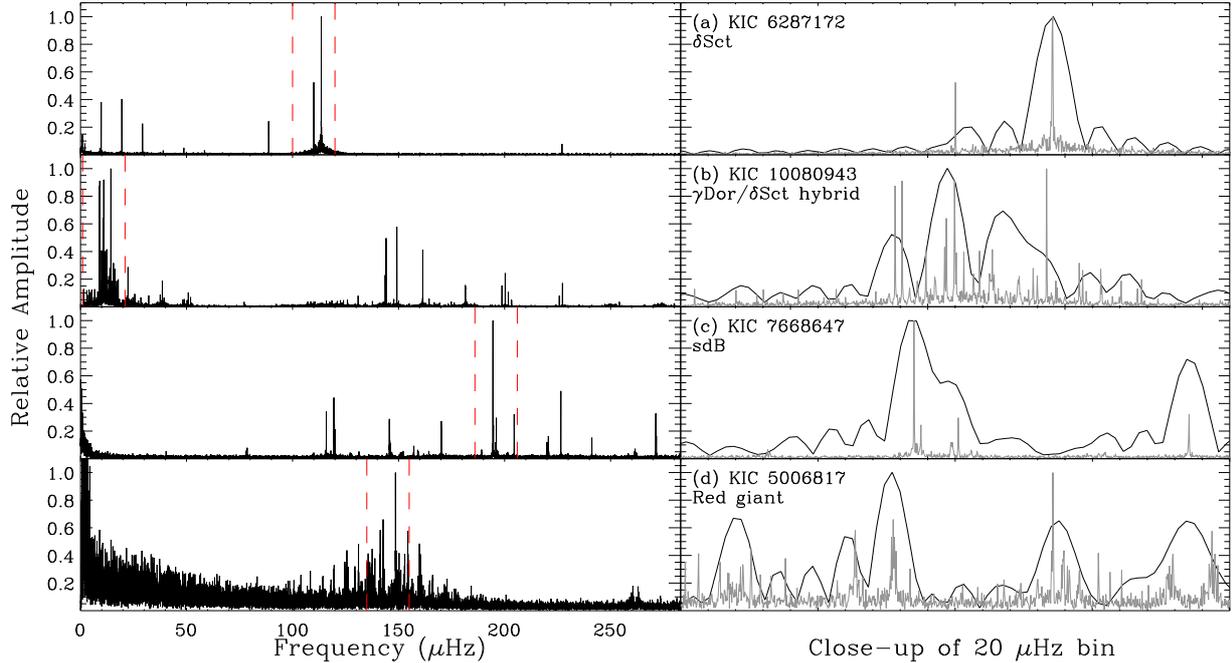}
	\caption{Examples of amplitude spectrum for each type of star analysed in this paper ordered from highest to lowest ${\rm S/N}$. The left column is the full spectrum calculated from the observed {\it Kepler} time series and the right column is a $20 \mu{\rm Hz}$-wide close-up of the region between the red dashed lines. The close-up plots show the spectra of the full time series (grey line) and a typical ten day sub-series (black line). The types of stars are as follows: (a) a $\delta$~Sct star with p-mode pulsations. (b) a $\gamma$~Dor/$\delta$~Sct hybrid star with low frequency g mode and high frequency p mode pulsations. (c) an sdBs with g mode pulsations. (d) a red giant branch star with solar-like mixed dipole modes oscillations. The amplitude of each plot has been normalised to the height of the strongest pulsating peak.}
	\label{fig:starex}
\end{figure*}

We looked for phase modulation {\it Kepler} data for in a variety of pulsating stars with known binaries. Examples are shown in Fig.~\ref{fig:starex}, and all have coherent modes with lifetimes longer than the four years of the observations. We considered two types of main-sequence pulsating stars near the instability strip. $\delta$~Scuti stars have high-amplitude and high-frequency p modes, which is why they were initially chosen when the PM method was developed. $\gamma$ Doradus stars have g mode pulsations at lower frequencies. In addition, we looked at $\delta$~Sct/$\gamma$ Dor, hybrids which have both $\delta$~Sct and $\gamma$~Dor pulsations. We also investigated red giant branch (RGB) and clump stars, where the coupling between pressure and gravity-dominated modes generates mixed dipole modes with long lifetimes~\citep[e.g.][]{dupret09}. Finally, we also considered two classes of compact evolved stars, the subdwarf~B (sdB) stars and white dwarfs, which both have coherent and modest-amplitude g modes~\citep[e.g.][]{reed11,greiss14} that could be suitable for detecting binary phase modulations.

We used {\it Kepler} light curves that had been reduced using the multiscale Maximum A Posteriori (msMAP) algorithm developed by~\citet{stumpe12}. This removes systematic trends, discontinuities, outliers, and artifacts. The light curves were then analysed using the PM method, as outlined in Section~\ref{sec:extract}.

Relating our simulations to the observed data requires knowledge of the projected light travel time across the orbit and the signal-to-noise ratios of the observed pulsations. For each star the strongest mode in the amplitude spectrum was selected for analysis and its frequency~$\nu_i$ and amplitude~$A_i$ were noted. The light travel time was calculated as described in section~\ref{sec:extract}. The signal-to-noise was calculated using Equation~\ref{equ:snamp} from the pulsation amplitude and the mean noise level. The results are plotted as symbols in Fig.~\ref{fig:ex}. The location of a star in Fig.~\ref{fig:ex} gives an estimate of the relative uncertainty of its maximum time delay for an individual mode. If the relative uncertainty is greater than 30\% it is unlikely to be detectable. Analysing multiple modes reduces the total relative uncertainty by approximately the square root of the number of modes.

Care must be taken when calculating the mean noise level, $\sigma_{\rm amp}$. For the highest amplitude $\delta$~Sct stars, the spectral window will dominate the amplitude spectrum \citep[e.g.][]{murphy13b}, even when the window function is ideal. To estimate the noise, we first pre-whitened the peak with the highest amplitude in the pulsation spectrum. The mean residual amplitude within $\pm 10$\% of the mode frequency was taken to be the mean noise level and used to calculate the signal-to-noise ratio. For most $\delta$~Sct stars, we note that this overestimates the noise because variance from other oscillation modes remains.

\section{Discussion}
\label{sec:disc}

Fig.~\ref{fig:ex} shows good agreement between the predicted detectability of binary stars and those with observed time-delay variations. This validates the use of the simulations as a way to determine a lower-bound of observable projected light travel time across the orbit for a given signal-to-noise and pulsation frequency. For example, a pulsation mode with $S/N = 100$ at $\nu = 210 \mu{\rm Hz}$ should typically show detectable phase modulation if the maximum time delay is greater than 30 seconds ($a_1 \sin{(i)}/c \cdot \nu = 30 \cdot 210 \cdot 10^{-6} = 0.0063$). 

If the primary star mass is known, a lower bound on the projected mass of the companion can be inferred. The lower bound on mass occurs when the orbital plane is perpendicular to the plane of the sky (i.e. $i=90^{\circ}$). However, the maximum time-delay for eccentric orbits depends on the argument of periapsis, $\varpi$. For the known binaries we examined, we assumed the eccentricity of the orbit has a negligible effect on the time-delays, i.e. the maximum time-delay is equivalent to $a_1 \sin{(i)}/c$ and does not affect the detectability. 

The $\delta$~Sct binaries in our sample, denoted by the red~squares and purple~triangles in Fig.~\ref{fig:ex}, were all detected using the PM method. Except for two, all lie above the maximum time-delay of noise (solid red line). These two were detected by analysing multiple modes (up to nine modes in total), which reduces the uncertainty of the time-delays by approximately the square root of the number of modes, whereas the simulations were calculated for single modes, only.

We found that $\gamma$~Dor pulsations are not suitable for the PM method because the frequencies of the g modes are too low. The examples shown in Fig.~\ref{fig:ex} are $\delta$~Sct/$\gamma$~Dor hybrid stars that only have detectable time-delays for the p-mode pulsations. The PM analysis of the g modes did not yield a detectable signal of binary motion. In general, the periods of the g modes are at least ten times greater than the p modes. This reduces the detectability of time-delay variations by an the same factor. Additionally, the density of modes in the Fourier transform of $\gamma$~Dor stars can be too high for them to be resolved with ten-day time intervals (see \citet{keen15} for an example); the time-delays are obscured by beating between other pulsation modes in the star. Binary $\delta$ Sct stars analysed by \citet{murphy14a} also show the effect of closely spaced beating modes on the time-delay spectrum.

From the analysis of RGB and clump stars we concluded that {\it Kepler} red giants will not have detectable time-delay variations. We inferred the maximum delays using 16 red giant binaries that have known orbital parameters reported by \citet{beck14} and \citet{gaulme14}. The ideal red giants are high-frequency, lower red-giant-branch (RGB) stars like the one shown in Fig.~\ref{fig:starex}d. We combined typical mass ratios of red giant binaries with a range of possible binary parameters to create a best-case scenario for the low luminosity RGB stars. The best red giants would exist in the dotted box in Fig.~\ref{fig:ex}. Therefore, we conclude that the signal-to-noise ratios of the coherent dipole modes, even in the best cases, are insufficient to detect time-delays caused by binarity. 

The companions of pulsating sdB stars are on the threshold of being detectable, as shown by the orange~diamonds in Fig.~\ref{fig:ex}. Two {\it Kepler} sdB stars in known binaries were analysed. The orbital periods are about 10 days, each with a white dwarf companion. Therefore, in an attempt to detect the phase modulation due to binarity, we took $N_{\rm sub}=500$ sub-series, which corresponded to a sub-series length of 2 days, which was required to sample the short orbit adequately. Naturally, decreasing the sub-series length increases the uncertainty of the time-delays, further reducing the detectability of the binarity of the sdBs. The PM method would succeed for sdB stars with longer orbital period companions, which would give a higher maximum time-delay. We were unable to detect binarity from the 19 pulsating sdB stars that have {\it Kepler} long cadence data, and which are not known to be binaries \citep[][]{silvotti14}. Previous work by \citet[][]{telting12,telting14} analysed these two sdB stars in a similar way by fitting the time delays to the pulsation modes. They were successful in extracting the maximum time delays by using {\it Kepler} short cadence data and using many tens of modes. Our analysis considers only a single mode, and is based on long-cadence data only.

We also considered white dwarfs, although we note that none of the pulsating white dwarfs in the {\it Kepler} field are in known binary systems. The pulsation frequency of white dwarfs is an order of magnitude higher than the other stars considered in this analysis. Therefore, {\it Kepler} short-cadence data (one minute sampling interval) were used in the analysis, which is not entirely comparable to our simulations in Fig.~\ref{fig:ex}. We infer from the work of \citet{hermes11} and \citet{greiss14} that the signal-to-noise ratios of white dwarf oscillations are similar to those of red giants. Note that the short oscillation periods of white dwarfs can cause the maximum time delay to exceed the pulsation period, in which case one must also account for phase wrapping of the binary-induced phase shifts. We concluded that the high frequency pulsations suit the PM method but, for the same reasons as for the sdBs, the smaller number of pulsating white dwarfs with {\it Kepler} data limits the chances of detecting binary systems.

\section{Conclusions}
\label{sec:conc}

We attempted to extend range of stars for which the phase modulation method can be applied, to include red giants, $\gamma$~Dor stars, white dwarfs, and subdwarf B stars. We explored the asteroseismic and orbital parameter space to find the detection limits. The results from our simulations show a relationship between the signal-to-noise ratio of the pulsation mode and the ability to detect binarity. To confirm this, we compared the results of the simulations with observed {\it Kepler} light curves. We saw a strong agreement in binary detectability between the observed and simulated data. Moreover, for $\delta$~Scts star with the highest signal-to-noise ratios, time-delay variations as low as a few seconds should be detectable. For a monoperiodic oscillator, this maximum time-delay corresponds to a companion mass on the order of $M \sin{(i)} = 10$ Jupiter masses, given the limits of the {\it Kepler} time series. This limit can be decreased for stars with pulstions above the Nyquist frequency or when analysing multiple modes, which can reduce the time-delay uncertainty by $\sqrt{N}$, where $N$ is the number of modes analysed. The limit for $\gamma$ Dor and sdBs is approximately 10 times more massive, due to the differences in mode frequency and signal-to-noise ratio. These companions would be very-low-mass stars that have periods of over one year. The limit for red giants is 100 times greater relative to the $\delta$ Sct stars, which is on the order of a solar mass companion. This optimistic case does not take into account the density of g modes, which cause heavy interference in the time-delay spectra, ultimately causing the red giants to be unrealistic candidates for the PM method.

We conclude that the PM method is best suited to searching for companions around $\delta$ Sct stars, where it should be possible to reach down to planetary masses.

\section*{Acknowledgements}
\label{sec:ack}

This research was supported by the Australian Research Council. Funding for the Stellar Astrophysics Centre is provided by the Danish National Research Foundation (grant agreement no.: DNRF106). The research is supported by the ASTERISK project (ASTERoseismic Investigations with SONG and Kepler) funded by the European Research Council (grant agreement no.: 267864).

\bibliographystyle{mnras}
\bibliography{ref}

\appendix

\label{lastpage}
\end{document}